# Topologically enhanced optical helicity density in the thermal near field of twisted bilayer van der Waals materials

Xiaohong Zhang,[1,*] Chiyu Yang,[2] Wenshan Cai,[1] and Zhuomin M. Zhang[2,†]

[1]*School of Electrical and Computer Engineering, Georgia Institute of Technology, Atlanta, Georgia, USA*
[2]*George W. Woodruff School of Mechanical Engineering, Georgia Institute of Technology, Atlanta, Georgia, USA*

**ABSTRACT**. Twisted van der Waals (vdW) bilayers can support tunable surface/hyperbolic phonon polariton (S/HPhP) depending on the interlayer twist angle. S/HPhPs can be thermally excited and significantly modify the thermal near field. A photonic topological phase transition occurs at a critical twist angle where the polariton dispersion switches from hyperbolic to elliptical. Because the twist angle governs the polariton modes, it is intrinsically linked to the optical helicity density (OHD) of the near-field thermal emission. In this work, a relationship between the OHD of near-field emission and the twist angle of bilayer twisted vdW materials is discovered and investigated. To evaluate the OHD, a 3×3 coherence matrix method is obtained from the fluctuation-dissipation theorem (FDT), which provides a complete description of the thermal electromagnetic field of the twisted bilayer, and a formalism for OHD based on the polarization matrix is employed. The topological transition angle (TTA) is determined by calculating the polariton dispersion relation of the vdW bilayer at different twist angles. A strong correlation between OHD and TTA is observed, which can be attributed to polariton canalization and confined group velocity, leading to enhanced polariton directionality. This study provides new insights into the analysis of angular momentum in near-field thermal radiation from twisted vdW structures.

## I. INTRODUCTION

Thermal radiation is traditionally regarded as a simple means of generating broadband, unpolarized, and uncollimated light, and therefore has been widely used in practical applications as mid or far infrared light sources [1]. However, additional optical components, such as filters or polarizers, are needed to improve the coherence of thermal radiation as required by some emerging advanced applications, e.g., polarized thermal imaging [2] and polarized infrared/terahertz spectroscopy [3]. Metamaterials provide an alternative route to producing partially coherent thermal radiation without external components by supporting resonant electromagnetic modes, including waveguide modes [4], surface phonon polariton (SPhP) [5–7], bound states in the continuum [8], or magnetic polaritons [9–12]. These capabilities have enabled applications in thermal photovoltaics, radiative cooling, and other advanced technologies. Among various metamaterials, twisted van der Waals (vdW) bilayers exhibit unique polaritonic properties through the hybridization of the iso-frequency contour (IFC) of each layer [13,14]. Analogous to the Lifshitz transition in condensed-matter physics [15] where the Fermi surface undergoes a change in its topology, the twisted vdW bilayer also undergoes a photonic topological transition when the twist angle exceeds a critical value, known as the topological transition angle (TTA). Across this transition, the IFC topology switches between hyperbolic and elliptical, leading to different optical responses [16]. Near the TTA, a highly confined and directional polariton mode called polariton canalization can be achieved through twisted vdW bilayers [13,14,17], graphene-hexagonal-boron-nitride (hBN) grating [18], or even a single flake of vdW material at a certain wavelength [19]. Specifically for twisted vdW bilayers, canalized polariton is associated with strong coherence of the thermal near field. Such canalized polariton modes have shown great potential for enhanced light-matter interactions [18], super-resolution imaging [20], and quantum materials [21]. Therefore, it is of great significance to study the relation between thermal near-field optical properties and photonic topological transition on the twisted vdW bilayers.

In this paper, angular momentum in the thermal near field of twisted vdW bilayers is analyzed. Nonzero angular momentum in the far field has been observed in a variety of metamaterials, which has been extensively realized through symmetry-breaking structures [8,22], twisted metasurfaces [7,23,24], or intrinsically nonreciprocal materials [25]. In most cases, spin angular momentum density (SAM) and optical helicity (chirality) density (OHD) are two different, but closely related, descriptions used for quantifying the angular momentum of light. Neither SAM nor OHD can completely account for the total angular momentum of an arbitrary light [26,27]. The SAM and OHD can be converted to each other for paraxial light (two-dimensional planar wavefront), but such a relation cannot be extended to three-dimensional electromagnetic fields, e.g., non-paraxial light or evanescent waves [28]. It has recently been theoretically proven that, for non-paraxial or in the near field, the SAM of the electromagnetic field thermally radiated by a bulk reciprocal material vanishes [25]. It should be emphasized that this statement only applies to structured light, because it has been shown many times that the SAM is nonzero for ordinary paraxial circularly polarized light (two-dimensional). However, the OHD in the thermal near field can remain nonzero and thus serves as a meaningful physical quantity to describe the near-field angular momentum. The interplay between near-field OHD and topological transition in twisted vdW bilayers has also shown important potential in near-field radiative heat transfer and near-field optics [29,30]. In this work, the OHD of multiple kinds of vdW bilayers, as well as their relationship with photonic topological phase transition, is investigated. An indirect approach based on the fluctuation-dissipation theorem (FDT) is implemented to calculate the electromagnetic coherence matrix and to obtain the OHD. The effects of twist angle and layer thickness are also studied, and a strong correlation between near-field helicity density and TTA is discovered, which may be explained by polariton canalization.

*xzhang926@gatech.edu

†zhuomin.zhang@me.gatech.edu

## II. THEORY

### A. Fluctuational electrodynamics

Thermal radiation can be understood as the collective emission of infinitesimally fluctuating dipoles inside the emitter. The behavior of these dipoles is governed by FDT, which was systematically formulated for thermal radiation by Rytov [31] in the 1950s and further extended to the near-field radiative heat transfer later by Polder and van Hove [32]. This framework provides a fundamental way to analyze thermally radiated electromagnetic fields from materials in both the near- and far-field regimes.

Starting with the general expression for magnetic vector potential $\mathbf{A}$ at the observation point $\mathbf{r}_1$ and the source point $\mathbf{r}'$:

$$\mathbf{A}(\mathbf{r}_1, \omega) = \mu_0 \int_V \bar{\bar{G}}(\mathbf{r}_1, \mathbf{r}', \omega) \mathbf{J}(\mathbf{r}', \omega) d^3\mathbf{r}', \quad (1)$$

where $\bar{\bar{G}}$ is the dyadic Green's function and $\mathbf{J}$ is the fluctuational current density. The corresponding electric field $\mathbf{E}$ and magnetic field $\mathbf{H}$ take the forms as follows:

$$\mathbf{E}(\mathbf{r}_1, \omega) = i\omega \mathbf{A}$$
$$= i\omega \mu_0 \int_V \bar{\bar{G}}(\mathbf{r}_1, \mathbf{r}', \omega) \mathbf{J}(\mathbf{r}', \omega) d^3\mathbf{r}' \quad (2)$$

$$\mathbf{H}(\mathbf{r}_1, \omega) = \nabla_{\mathbf{r}_1} \times \mathbf{A}$$
$$= \nabla_{\mathbf{r}_1} \times \mu_0 \int_V \bar{\bar{G}}(\mathbf{r}_1, \mathbf{r}', \omega) \mathbf{J}(\mathbf{r}', \omega) d^3\mathbf{r}', \quad (3)$$

where $i$ is the unit imaginary number, $\omega$ is the angular frequency, and $\mu_0$ is the vacuum permeability. The operator $\nabla_{\mathbf{r}_i} \times$ indicates the curl operator acting on $\mathbf{r}_i$. From Eq. (1), one can derive the cross-spectral tensor, which will be used for calculating the OHD in the next section, by taking the cross-correlation between $\mathbf{E}(\mathbf{r}_1)$ and $\mathbf{H}(\mathbf{r}_2)$ at two locations $\mathbf{r}_1$ and $\mathbf{r}_2$, as in Eq. (4) and Eq. (5).

$$\langle \mathbf{E}_{\text{ge}}(\mathbf{r}_1, \omega) \otimes \mathbf{H}_{\text{ge}}^*(\mathbf{r}_2, \omega) \rangle$$
$$= \frac{i\omega}{\mu_0} \langle \mathbf{A}(\mathbf{r}_1, \omega) \otimes \mathbf{A}^*(\mathbf{r}_2, \omega) \rangle (\nabla_{\mathbf{r}_2} \times)^{\text{T}}, \quad (4)$$

$$\langle \mathbf{A}(\mathbf{r}_1, \omega) \otimes \mathbf{A}^*(\mathbf{r}_2, \omega) \rangle$$
$$= \mu_0^2 \iint_V \bar{\bar{G}}(\mathbf{r}_1, \mathbf{r}', \omega) \langle \mathbf{J}(\mathbf{r}', \omega) \otimes \mathbf{J}^*(\mathbf{r}'', \omega) \rangle$$
$$\times \bar{\bar{G}}^\dagger(\mathbf{r}_2, \mathbf{r}'', \omega) d^3\mathbf{r}' d^3\mathbf{r}'', \quad (5)$$

where the subscript "ge" stands for "global thermal equilibrium" for the emitter, and $\bar{\bar{A}} \otimes \bar{\bar{B}} = \bar{\bar{A}} \bar{\bar{B}}^{\text{T}}$ is the outer product, where the superscript "T" is transpose. The first kind of FDT models the correlation of $\mathbf{J}$ between two locations:

$$\langle \mathbf{J}(\mathbf{r}', \omega) \otimes \mathbf{J}^*(\mathbf{r}'', \omega) \rangle = \frac{4}{\pi} \omega \varepsilon_0 \Theta(\omega, T)$$
$$\times \frac{\bar{\bar{\varepsilon}}(\omega) - \bar{\bar{\varepsilon}}^\dagger(\omega)}{2i} \delta(\mathbf{r}' - \mathbf{r}''), \quad (6)$$

where $\Theta(\omega, T) = \hbar\omega/2 + \hbar\omega/(e^{\hbar\omega/k_B T} - 1)$ is the mean energy of Planck's oscillator, with $k_B$ the Boltzmann constant, $\hbar$ the reduced Planck constant, and $T$ the temperature. $\varepsilon_0$ and $\bar{\bar{\varepsilon}}$ are the permittivity of the vacuum and the relative permittivity tensor of the medium. $\dagger$ and $\delta$ denote the Hermitian conjugate and Dirac's delta, respectively. With Eq. (6), Eq. (5) can be further written as:

$$\langle \mathbf{A}(\mathbf{r}_1, \omega) \otimes \mathbf{A}^*(\mathbf{r}_2, \omega) \rangle = \frac{4}{\pi} \frac{\mu_0 \omega \Theta(\omega, T)}{c^2}$$
$$\times \int_V \bar{\bar{G}}(\mathbf{r}_1, \mathbf{r}', \omega) \frac{\bar{\bar{\varepsilon}}(\omega) - \bar{\bar{\varepsilon}}^\dagger(\omega)}{2i} \bar{\bar{G}}^\dagger(\mathbf{r}_2, \mathbf{r}', \omega) d^3\mathbf{r}', \quad (7)$$

where $c = (\varepsilon_0 \mu_0)^{-1/2}$ is the speed of light in vacuum. An important identity can be imposed to simplify the integral with Green's function [33]:

$$\frac{\omega^2}{c^2}\int_V \bar{\bar{G}}(\mathbf{r}_1,\mathbf{r}',\omega)\frac{\bar{\bar{\varepsilon}}(\omega)-\bar{\bar{\varepsilon}}^\dagger(\omega)}{2i}\bar{\bar{G}}^\dagger(\mathbf{r}_2,\mathbf{r}',\omega)d^3\mathbf{r}'$$
$$=\frac{\bar{\bar{G}}(\mathbf{r}_1,\mathbf{r}_2,\omega)-\bar{\bar{G}}^\dagger(\mathbf{r}_2,\mathbf{r}_1,\omega)}{2i}, \quad (8)$$

and therefore, Eq. (7) becomes:

$$\langle \mathbf{A}(\mathbf{r}_1,\omega)\otimes\mathbf{A}^*(\mathbf{r}_2,\omega)\rangle = \frac{4}{\pi}\mu_0\frac{\Theta(\omega,T)}{\omega}$$
$$\times\frac{\bar{\bar{G}}(\mathbf{r}_1,\mathbf{r}_2,\omega)-\bar{\bar{G}}^\dagger(\mathbf{r}_2,\mathbf{r}_1,\omega)}{2i}. \quad (9)$$

Denote $\bar{\bar{G}}(\mathbf{r}_i,\mathbf{r}_j,\omega)$ as $\bar{\bar{G}}_{ij}$ for simplicity. The dyadic Green's function can be expressed in the k-space in terms of its spatial Fourier transform $\bar{\bar{g}}$:

$$\bar{\bar{G}}_{12} = \frac{1}{(2\pi)^2}\iint \bar{\bar{g}}(z_1,z_2,\mathbf{k}_\parallel,\omega)e^{i\mathbf{k}_\parallel\cdot(\mathbf{R}_1-\mathbf{R}_2)}d^2\mathbf{k}_\parallel, \quad (10)$$

where $\mathbf{R}_i = x\hat{\mathbf{x}} + y\hat{\mathbf{y}}$ and $\mathbf{k}_\parallel = k_x\hat{\mathbf{x}} + k_y\hat{\mathbf{y}}$. For any stacked homogeneous planar medium, $\bar{\bar{g}}$ has an explicit form (See Appendix A) in terms of the reflection coefficients at the top interface. Hence, the explicit form of Eq. (4) is:

$$\langle \mathbf{E}_{\text{ge}}(\mathbf{r}_1,\omega)\otimes\mathbf{H}_{\text{ge}}^*(\mathbf{r}_2,\omega)\rangle$$
$$= \frac{4i}{\pi}\Theta(\omega,T)\frac{\bar{\bar{G}}_{12}-\bar{\bar{G}}_{21}^\dagger}{2i}(\nabla_{\mathbf{r}_2}\times)^{\text{T}}, \quad (11)$$

where the curl operator also takes a simple matrix form, whose derivation is shown in Appendix B. Since $\bar{\bar{g}}$ is the k-space Fourier transform of $\bar{\bar{G}}$, the resulting correlation function of the electric field naturally takes evanescent waves and other near-field phenomena into account.

When global thermal equilibrium is satisfied, the total field cross-correlation consists of two parts [34,35]: (1) The native radiated field by the medium itself; and (2) The radiation from the surrounding environment, which can be modeled as blackbody radiation at the same temperature, emitted in vacuum from $r = \infty$. The expression for the radiation from the environment, $\mathbf{E}_{\text{vac}}$ and $\mathbf{H}_{\text{vac}}$, are listed in Appendix A. Therefore, the native emitted field of the medium itself and the correlation between each field component, also known as the 3×3 coherence matrix, $\bar{\bar{W}}$, for $\mathbf{E}$ and $\mathbf{H}$, can be obtained by subtracting the contribution of the background with the total field correlation:

$$\bar{\bar{W}} = \langle \mathbf{E}(\mathbf{r}_1,\omega)\otimes\mathbf{H}^*(\mathbf{r}_2,\omega)\rangle$$
$$= \langle \mathbf{E}_{\text{ge}}(\mathbf{r}_1,\omega)\otimes\mathbf{H}_{\text{ge}}^*(\mathbf{r}_2,\omega)\rangle \quad (12)$$
$$-\langle \mathbf{E}_{\text{vac}}(\mathbf{r}_1,\omega)\otimes\mathbf{H}_{\text{vac}}^*(\mathbf{r}_2,\omega)\rangle,$$

whose closed form is:

$$\bar{\bar{W}} = \begin{bmatrix} \langle E_x H_x^\dagger\rangle & \langle E_x H_y^\dagger\rangle & \langle E_x H_z^\dagger\rangle \\ \langle E_y H_x^\dagger\rangle & \langle E_y H_y^\dagger\rangle & \langle E_y H_z^\dagger\rangle \\ \langle E_z H_x^\dagger\rangle & \langle E_z H_y^\dagger\rangle & \langle E_z H_z^\dagger\rangle \end{bmatrix}. \quad (13)$$

### B. Three-dimensional optical helicity density and chirality

The angular momentum of light was discovered a long time ago. Early in 1936, Beth [36] first experimentally measured non-zero time-averaged angular momentum in paraxial light. Light with chirality, or structured light, can be used for chemical and biological applications, such as enantiomer detection and discrimination [37]. The angular momentum of light generated by metamaterials has attracted growing attention in recent years. Extensive theoretical efforts have been devoted to modeling the angular momentum of light. Closely related to the circular polarization of light, two physical quantities describe the "intrinsic angular momentum" of a photon: (1) spin angular momentum (SAM) density and (2) optical helicity (chirality) density (OHD). SAM is a pseudovector that partially contributes to the total angular momentum, while the remaining is from orbital angular momentum [27]. For paraxial light, SAM and OHD are proportional to each other and can be directly obtained from the Stokes parameter $S_3$ for circular polarization. The optical helicity (chirality), defined analogously to the magnetic field-line helicity in plasma physics, quantifies the winding and curling of the field lines and is thus a pseudoscalar, distinct from SAM [26,27]. More importantly, the difference between SAM and OHD becomes significant for near-field thermal radiation, because in the thermal near field, the radiation is almost always non-paraxial, i.e., a three-dimensional electromagnetic field. It has been theoretically proven that, under such conditions, the SAM is always zero for structures that obey reciprocity [25]. The zero SAM can also be interpreted differently, since for a three-dimensional field, SAM can be projected onto each dimension, and each corresponds to one of the three-dimensional Stokes parameters representing circular polarizations [38]. In the near field, especially when polariton modes are excited, the electric field is dominated by longitudinal components as well as strong phase-incoherent radiation from different directions and thus cannot sustain circular polarization. Therefore, the SAM stays at zero for non-paraxial thermal radiation from reciprocal materials. In contrast, OHD is defined through the curl and topology of the electromagnetic field lines and therefore does not necessarily have a fixed relation to circular polarization in this case. Hence, as will be presented in Sec. III, the OHD in the near field can remain nonzero even for reciprocal materials.

The OHD, $h$, has a general form as follows [28,39–41]:

$$h = \frac{1}{2}\varepsilon_0 c(\mathbf{A} \cdot \mathbf{B} - \mathbf{C} \cdot \mathbf{E}), \quad (14)$$

where $\mathbf{A}$ and $\mathbf{C}$ are the transverse components of the magnetic and the electric vector potentials. The OHD is a conserved and gauge-invariant quantity when no charge is present [39]. The only transverse components of the vector potentials contribute to the OHD [28,40]. A simplified form can be written for paraxial light produced by a non-magnetic material [28]:

$$h = -\frac{1}{2\omega c}\text{Im}(\mathbf{E}^* \cdot \mathbf{H}). \quad (15)$$

For elliptically polarized light, $\mathbf{E}$ and $\mathbf{H}$ are complex vectors, and thus $h$ does not vanish. In vacuum, for example, $h$ reaches its maximum absolute value for circular polarization, $h = \pm\varepsilon_0/\omega$, while it vanishes for linear polarization. For general non-paraxial thermal emission with three-dimensional field distribution (e.g., near field or evanescent wave), the OHD takes the following revised form [28]:

$$h = -\frac{1}{2\omega c}\text{Im}\left(\text{Tr}(\bar{\bar{W}}^* - \bar{\bar{W}}^\text{T})\right), \quad (16)$$

where $\bar{\bar{W}}$ is the 3×3 coherence matrix obtained through the FDT discussed in Sec. II.A, and "Tr" stands for trace. It is worth pointing out that chirality and OHD are interchangeable qualities and are simply linked by the following relation [28,42]:

$$\chi = \omega k_0 h = -\frac{k_0}{2c}\text{Im}\left(\text{Tr}(\bar{\bar{W}}^* - \bar{\bar{W}}^\text{T})\right), \quad (17)$$

where $k_0$ is the free-space wavenumber. For the rest of this paper, only OHD is used for simplicity.

### III. RESULTS AND DISCUSSION

In this section, the calculation results for the OHD near (1) a bilayer of α-molybdenum trioxide (α-MoO$_3$) and (2) a bilayer of hexagonal boron nitride (hBN) are presented. These two vdW materials are selected as representative examples to demonstrate the generality of the association between the OHD peak and the TTA in twisted vdW bilayers. The schematic of the twisted vdW bilayer and the definitions of the geometric parameters and coordinates shared by both cases are illustrated in Fig 1(a), along with the dielectric functions of α-MoO$_3$ in Fig. 1(b) and that of hBN in Fig. 1(c). The effects of twist angle and layer thickness are investigated, especially near the TTA. Different wavelengths are considered, and the shifting of the OHD peak, as well as the TTA, with respect to the wavelength, is also discussed.

#### A. Bilayer α-MoO$_3$

As a biaxial crystal, α-MoO$_3$ supports hyperbolic phonon polaritons within three Reststrahlen bands, as shown in Fig. 1(b) [43,44]. The in-plane hyperbolicity of its permittivity tensor originates from the metal-like behavior ($\varepsilon < 0$) along one principal axis and dielectric-like behavior along the other ($\varepsilon > 0$). When stacking one with the other to form a twisted bilayer, the α-MoO$_3$ bilayer can support different hybridized polariton modes. The resulting in-plane polariton dispersion, i.e., IFCs in the k-space, can transition between hyperbolic and elliptical shapes as the twist angle varies. Notably, at TTA, the eccentricity of the hyperbolas reaches its maximum among all twist configurations, i.e., a pair of nearly parallel lines, marking the boundary between hyperbolic and elliptical dispersion. This phenomenon is referred to as polariton canalization. Hu et al. [13] have experimentally verified this photonic topological transition by probing the polariton modes launched from an artificial point defect on a twisted α-MoO$_3$ bilayer using scanning near-field optical microscopy. Their work also verifies the existence of polariton canalization, in which the group velocity of hyperbolic phonon polaritons (HPhPs) is restricted to a single direction and its opposite, thereby strongly confining the propagation of HPhP.

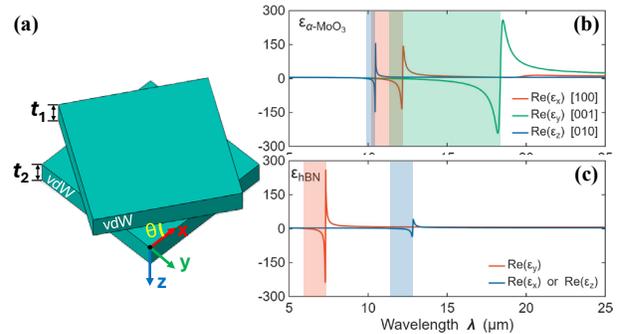

FIG. 1. (a) Schematic of a twisted vdW bilayer. Both layers are assumed to be infinitely large in the $xy$-plane and embedded in vacuum. The two layers share the same crystal orientation before twisting. The coordinate system has the $z$-direction pointing downward. The twist angle is defined with respect to the $x$-axis. $t_1$ and $t_2$ denote the thickness of the top and the bottom layer, respectively. (b) and (c) Anisotropic dielectric functions of α-MoO$_3$ and hBN, respectively. The Reststrahlen bands are highlighted by color-shaded regions. For hBN, the optical axis of each layer is initially aligned along the $y$-axis at $\theta = 0$.

Thermal radiation provides broadband excitation, producing not only propagating waves with in-plane wavenumber $k_\parallel < k_0$ but also evanescent waves or surface waves with $k_\parallel > k_0$. As longitudinal waves, SPhP or HPhP strongly affect the degree of polarization and other optical properties in the near field and consequently affect the OHD of the field [45].

The geometric arrangement of the α-MoO$_3$ bilayer is shown in Fig. 1(a). Each layer has its [010] crystal axis aligned along the z-axis. Both layers are assumed to be infinitely large in the xy-plane. The distance between the observation point and the top surface is denoted by $d$. To evaluate the OHD of this structure in the region $z < 0$, the method based on the optical transfer matrix is used to calculate the Fresnel coefficients of the structure [46]. FDT is then applied to obtain the coherence matrix following the method discussed in Sec. II.A, and the OHD is subsequently obtained according to Eq. (16).

The results in Fig. 2, at two different wavelengths $\lambda$ = 11 μm and 11.36 μm, show that the OHD increases with the twist angle, reaches a maximum near the TTA, and then decreases as the twist angle increases further. Fig. 3 shows the polariton dispersion of the bilayer and the IFC of each layer. These result together indicates a clear correlation between the OHD maximum and the TTA. Two observations are drawn for twisted α-MoO$_3$ bilayers: (1) The OHD is inversely related to the absolute deviation of the twist angle from the TTA at a given wavelength; the closer the system is to the topological transition, the larger the OHD. (2) For $\theta \in [0, 90°]$, only one maximum of the OHD occurs at TTA.

Figs. 2(a) to (c) show the OHD at $\lambda = 11$ μm. In Fig. 2(a), the OHD maximum occurs at around 61°, while the angle increases as the distance to the top surface, $d$, increases, as shown in Fig. 2(b). In Fig. 2(c), the OHD decays rapidly with increasing $d$, indicating the enhanced OHD is a proximal effect associated with the near-field radiation. In Fig. 3(a) and (b), the TM reflection coefficient $r_{pp}$ at $\lambda = 11$ μm is calculated, and its imaginary part is plotted to show mode dispersion and energy distribution. At $\lambda$ = 11 μm within the second Reststrahlen band, $\varepsilon_x = -3.69$ and $\varepsilon_y = 1.18$, implying a hyperbolic IFC. The IFCs of the top and bottom layers intersect at two points when $\theta \leq 60°$ and at four points when $\theta > 60°$. Once polariton canalization is achieved near the TTA, $\theta = 60°$, the directionality of the HPhP reaches its maximum. It can be observed that in Figs. 2(a) and (b), the OHD peak also occurs close to 60°, suggesting that there is a possible link between the OHD peak and the TTA. Such a correlation can be explained by the confined group velocity of the polariton modes when close to the TTA, since the direction of group velocity is normal to the IFCs. Only for this specific combination of wavelength and twist angle can the group velocity be confined to a particular direction, thereby enhancing the OHD.

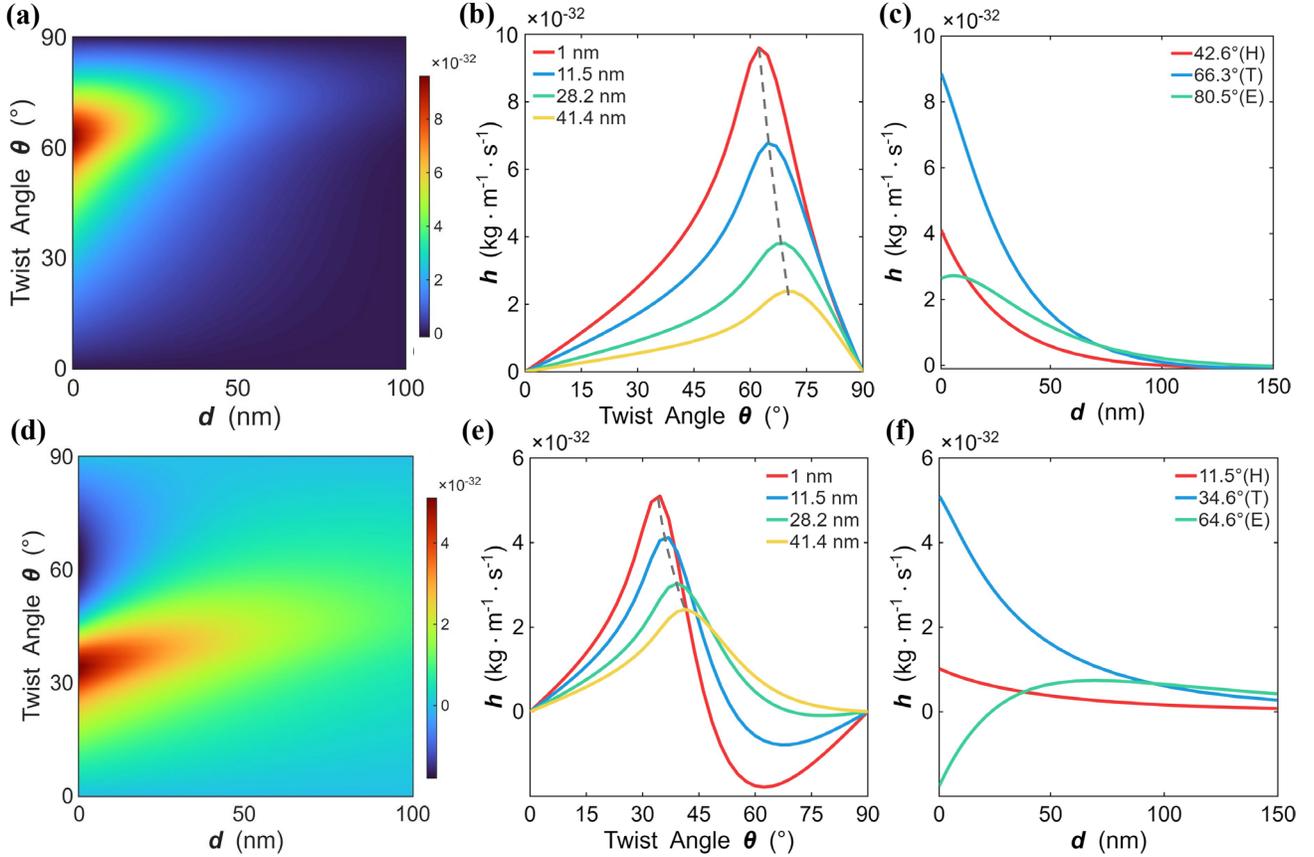

FIG. 2. OHD($h$) at $\lambda = 11$ μm (panels (a)–(c)) and 11.36 μm (panels (d)–(f)). The structure temperature is 300 K. Both layers are 100 nm. (a) and (d) OHD of a twisted α-MoO$_3$ bilayer as a function of twist angle and distance from the top surface. (b) and (e) Cross-sectional profile at selected distance $d$ from panel (a) and (d), respectively. (c) and (f) $h$ versus distance from the surface. The negative $h$ value means the field line changes its spiral direction.

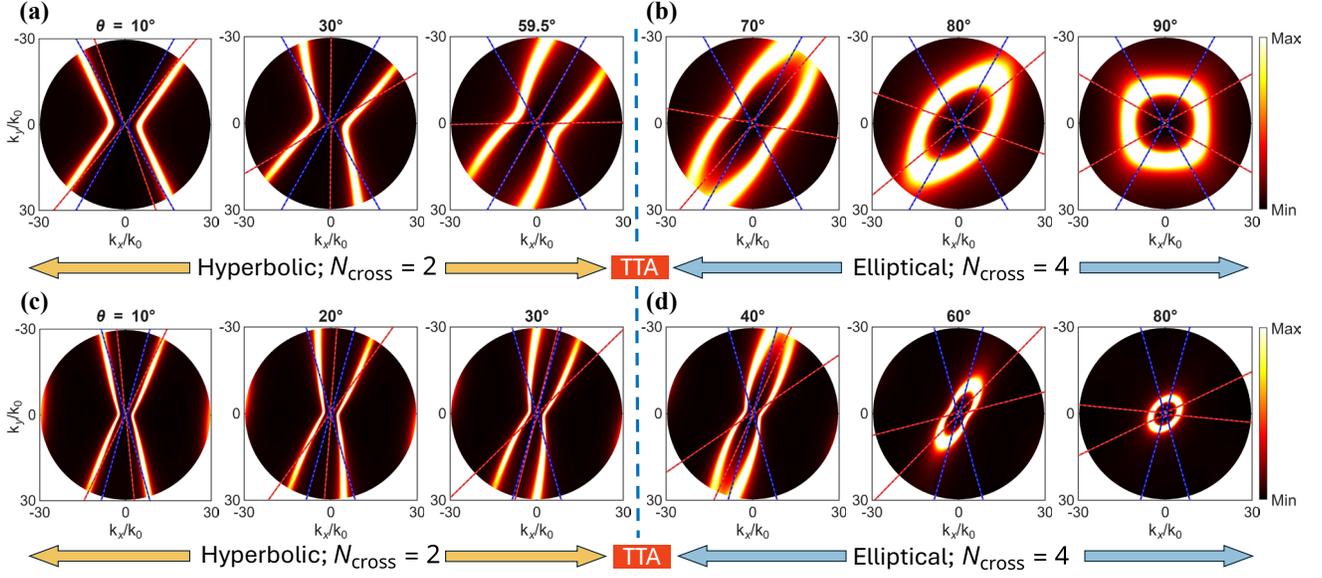

FIG. 3. Imaginary part of TM reflection coefficient, Im($r_{pp}$), and hyperbolic IFCs of the twisted bilayer a-MoO$_3$ at $\lambda$ = 11 μm ((a) and (b)) and 11.36 μm ((c) and (d)). The red and blue hyperbolas correspond to the IFCs of the top and the bottom layer, respectively. In each plot, $k_y$ is inverted to follow the same coordinate definition in Fig. 1(a). The grey dotted lines indicate the asymptotes of the theoretical hyperbolic IFCs of each layer. The dark blue dotted line divides the region into the hyperbolic and the elliptical region. The total dispersion is hyperbolic for θ ≤ TTA and elliptical for θ > TTA. Different IFC topologies are categorized by the number of intersections of individual IFCs, $N_{cross}$. For the hyperbolic regime, $N_{cross}$ = 2, and for the elliptical regime, $N_{cross}$ = 4. Only TM is considered here because of the excitation condition of the HPhP.

At a different wavelength $\lambda$ = 11.36 μm, with $\varepsilon_x = -8.37$ and $\varepsilon_y = 0.66$, the same correlation is observed. In Figs. 2(d) to (f), the OHD attains its maximum near 32°, in agreement with the prediction from the IFC in Figs. 3(c) and (d). It is worth noting that the TTA is wavelength-dependent, and so is the OHD maximum. As shown in Fig. 4, the TTA shifts to larger angles as the wavelength decreases and vice versa. Also, the OHD maximum is distance dependent. For the two wavelengths considered in Figs. 2(b) and (d), as the observation point moves away from the surface, the peak position shifts towards larger twist angles.

To examine the generality of the previous observation, the relation between the theoretical TTA and the twist angle where the OHD reaches its maximum is investigated for a broader continuous wavelength range (as in Fig. 4). The theoretical value of TTA can be estimated from the opening angle of the hyperbola given by [13]:

$$\theta_{TTA} = \text{atan}\sqrt{\frac{-\text{Re}(\varepsilon_x)}{\text{Re}(\varepsilon_y)}}. \quad (18)$$

Note that this estimation of TTA uses only the real part of the dielectric function and treats each layer as independent; therefore, it does not consider the interlayer hybridization of IFCs. The actual TTA depends on both layers and is affected by the layer thickness, especially for thin layers, and the distance to the surface. Thus, the slight difference between the two curves in Fig. 4 can be explained. Nevertheless, it is clear that the maximum OHD always occurs close to the topological transition of the IFC, and they follow the exact same trend when the wavelength varies.

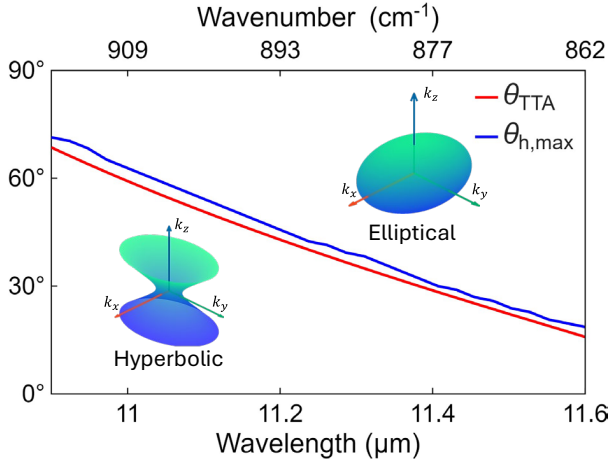

FIG. 4. Angles corresponding to the maximum OHD ($\theta_{h,\max}$; evaluated at $d = 1$ nm) and TTA ($\theta_{\mathrm{TTA}}$) as functions of wavelength. The same trend shared by both curves shows a strong correlation between the OHD and the TTA. The slight upward shift of $\theta_{h,\max}$ is due to the nonzero distance between the surface of the bilayer and the observation point. The transition between the hyperbolic and the elliptical region occurs upon crossing the boundary defined by the red curve.

To examine the influence of layer thickness on OHD, additional calculations were performed by independently varying the thicknesses of the top and bottom layers of the twisted bilayer α-MoO$_3$. The distance between the observation point and the top surface is set to 1 nm. The results are shown in Fig. 5, suggesting that the top layer has a strong impact on the OHD when its thickness $t_1$ is smaller than 20 nm. In contrast, the bottom-layer thickness shows a monotonic influence on the OHD during the initial increase in thickness. In both cases, the topological transition diminishes at small thicknesses, making the OHD more sensitive to thickness variations in that region. The OHD peak occurs when each layer is sufficiently thick, and it is consistently associated with the TTA, regardless of thickness, implying that it is topologically protected. Note that a similar observation on TTA is reported in [13]. As the thickness increases, its impact becomes weaker, and the OHD gradually decreases. In this region, each layer behaves more like a bulk material, and the coupling between layers is reduced due to increased loss.

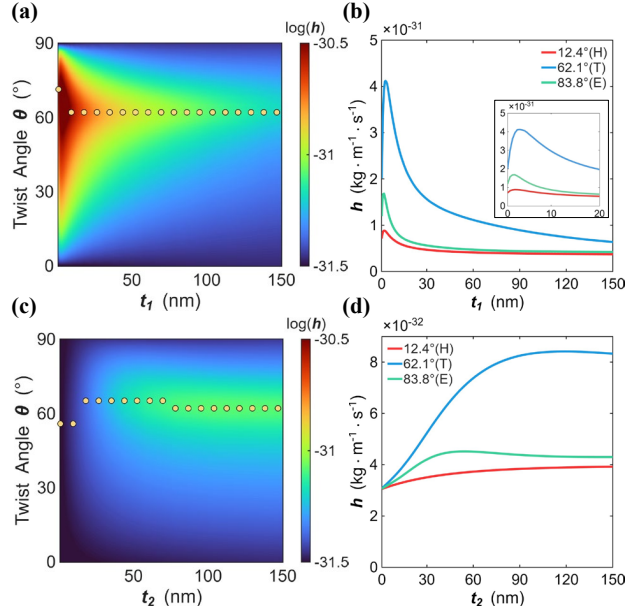

FIG. 5. The effect of layer thickness on OHD of the twisted α-MoO$_3$ bilayer at 300 K, with distance $d = 1$ nm. (a) and (b) OHD at different twist angles and different top layer thicknesses $t_1$, with $t_2 = 100$ nm. Inset of (b) shows the decrease in OHD when the thickness approaches zero. (c) and (d) OHD at different twist angles and different bottom layer thicknesses $t_2$, with $t_1 = 100$ nm. The yellow dots mark the maximum OHD for corresponding layer thicknesses.

### B. Bilayer hBN

As a classical vdW material, hBN consists of a two-dimensional hexagonal lattice analogous to graphene. Although hBN is intrinsically uniaxial, an in-plane hyperbolic permittivity tensor can be realized by orienting its optical axis in the $xy$-plane, leading to the dielectric function shown in Fig. 1(c) (data from Ref. [47]). HPhP modes are supported around $\lambda$ from 6.0-7.2 μm and from 12-13 μm, respectively, which offers great potential for mid-infrared applications, including infrared sensing and imaging, chemical detection, near-field radiative heat transfer, and quantum emitters [48–51].

Consider a twisted bilayer hBN structure, with the same arrangement in Fig. 1(a) and $t_1 = t_2 = 100$ nm. The optical axes are set in the $xy$-plane. The same procedure described in Sec. II is applied using the permittivity values shown in Fig. 1(c). The calculation results shown in Fig. 6 reveal a highly resembled behavior between OHD and twist angle, as in the case of α-MoO$_3$. An OHD maximum occurs near the structure's TTA, as shown in Figs. 6(a) and (b), consistent with the estimation based on the IFCs as shown in Fig. 6(d) and (e). Although the actual TTA of the bilayer is slightly ($< 5°$) different than the estimated TTA ($\theta_{\mathrm{TTA}}$) using the isolated IFC method, due to the thickness effect and interlayer coupling aforementioned, the link between the maximum OHD and TTA still holds. The cross-sectional profile plot along different twist angles, shown in Fig. 6(b), exhibits decay trends similar to those observed in the $\alpha$-MoO$_3$ bilayer. In addition, the same shifting of the OHD maximum towards larger angles (Fig. 6(b)) and fast decay of the peak (Fig. 6(c)) occurs as the observation point moves away from the surface.

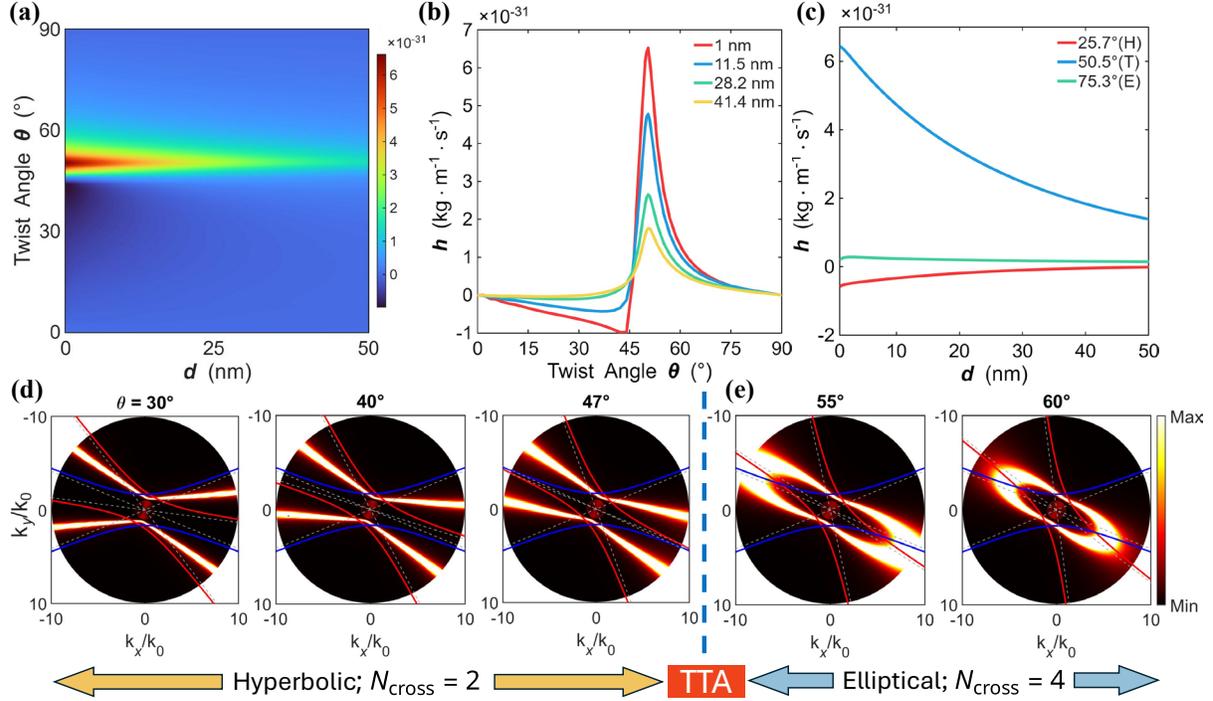

FIG. 6. OHD($h$) and IFCs of twisted hBN at $\lambda = 7$ μm. The temperature is 300 K, and both layers are 100 nm thick. (a) OHD of twisted bilayer hBN versus twist angle, $\theta$, and distance to the top surface, $d$. (b) Cross-sectional plot along different distances in (a). (c) Decay of the OHD along distance to the surface, obtained from cross-section along different angles in (a). (d) and (e) The imaginary part of the Fresnel coefficient for the TM wave, $\mathrm{Im}(r_{pp})$, and IFCs of the top (red) and bottom (blue) layer. Grey dotted lines denote the asymptotes of the hyperbolas, and the dark blue dotted line denote the boundary of topological transition of the IFC.

The layer thickness plays a vital role in OHD. A study on the influence of layer thickness at $d = 10$ nm was carried out, whose result is shown in Fig. 7. At very small thickness, the OHD peak cannot be observed, and once enough thickness is reached, the correlation between OHD maximum and TTA emerges and maintains for even larger thickness, showing the topologically protected nature. When the thickness of either layer exceeds around 90 nm, the OHD starts to decrease, but the maximum OHD still occurs at TTA. Along with the results of $\alpha$-MoO$_3$, it can be generalized that the topologically enhanced OHD can be realized in twisted bilayers made of different vdW materials with different thicknesses.

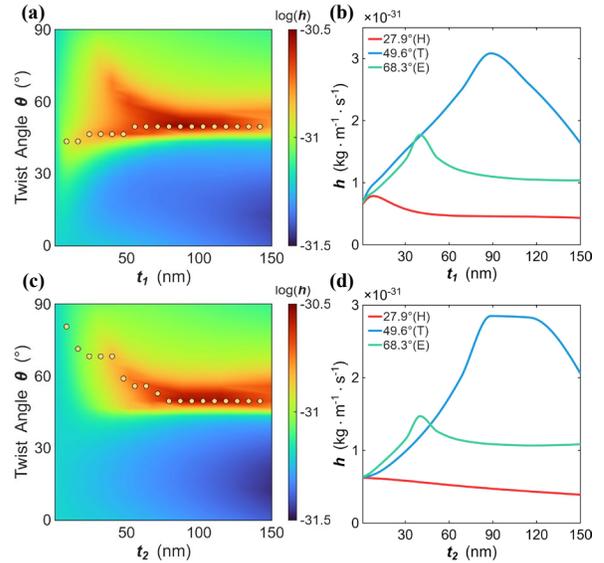

FIG. 7. The effect of layer thickness on OHD of the twisted hBN bilayer at 300 K ($d = 10$ nm). (a) and (b) OHD at different twist angles and different top layer thicknesses $t_1$, with $t_2 = 100$ nm. (c) and (d) OHD at different twist angle and different bottom layer thickness $t_2$, with $t_2 = 100$ nm. The yellow dots mark the maximum OHD for specific layer thicknesses.

## IV. CONCLUSIONS

The electromagnetic coherence matrix for thermal radiation from the twisted bilayer structure is obtained using fluctuational electrodynamics, from which the OHD of the three-dimensional thermal near field is calculated accordingly. A correlation between maximum OHD and TTA of two kinds of twisted bilayer vdW material, α-MoO$_3$ and hBN, is identified. The TTA is estimated through an independent-layer approximation, which also effectively captures the photonic topological transition of IFCs. The exact TTA is then determined by sweeping the twist angle and computing the corresponding IFCs. The results indicate that enhanced OHD is expected in hyperbolic materials at wavelengths where a topological transition of the IFC occurs. It is found that the OHD for non-paraxial thermal radiation has a proximity effect and decays rapidly with increasing distance from the emitter, indicating it is a near-field phenomenon. These findings contribute to the understanding of angular-momentum in near-field thermal radiation and may facilitate future studies in near-field optics, topological photonics, and radiative energy transfer.

## ACKNOWLEDGMENTS

This work was supported by the Department of Energy (DE-SC0018369) for X.Z. and Z.M.Z. and the National Science Foundation (DMR-2323909) for W.C.

## Appendix A: Dyadic Green's function and vacuum contribution

Here, the formulas used for deriving the dyadic Green's function are listed in detail. Continuing with the Fourier transform in Eq. (10), the spectral Green's function $\bar{\bar{g}}_{12}$ ($z_2 < z_1 < 0$) is given exactly by Eq. (A1) [23,25]:

$$\bar{\bar{g}}_{12}(\mathbf{k}_\parallel, \omega) = \underbrace{\frac{i}{2k_{0z}} e^{ik_{0z}(z_1-z_2)}[\hat{\mathbf{s}}_+\hat{\mathbf{s}}_+^{\mathrm{T}} + \hat{\mathbf{p}}_+\hat{\mathbf{p}}_+^{\mathrm{T}}]}_{\text{Vacuum contribution}} + \underbrace{\frac{i}{2k_{0z}} e^{ik_{0z}(-z_2-z_1)}[r_{ss}\hat{\mathbf{s}}_-\hat{\mathbf{s}}_+^{\mathrm{T}} + r_{sp}\hat{\mathbf{p}}_-\hat{\mathbf{s}}_+^{\mathrm{T}} + r_{pp}\hat{\mathbf{p}}_-\hat{\mathbf{p}}_+^{\mathrm{T}} + r_{ps}\hat{\mathbf{s}}_-\hat{\mathbf{p}}_+^{\mathrm{T}}]}_{\text{Structure response}} \quad (A1)$$

The first term of the summation corresponds to the vacuum contribution, while the second term corresponds to the desired structure's response. Note that, if $z_2 > z_1$, the exponential terms in the expression need to be modified, see Ref. [25]. For calculations in this paper, $z_1$ always equal to $z_2$ to get the autocorrelation of the electromagnetic field at an observation point. Similarly,

$$\bar{\bar{g}}_{21}(\mathbf{k}_\parallel, \omega) = \frac{i}{2k_{0z}} e^{ik_{0z}(z_1-z_2)}[\hat{\mathbf{s}}_+\hat{\mathbf{s}}_+^{\mathrm{T}} + \hat{\mathbf{p}}_+\hat{\mathbf{p}}_+^{\mathrm{T}}]$$
$$+ \frac{i}{2k_{0z}} e^{ik_{0z}(-z_2-z_1)}[r_{ss}\hat{\mathbf{s}}_-\hat{\mathbf{s}}_+^{\mathrm{T}}$$
$$+ r_{sp}\hat{\mathbf{p}}_-\hat{\mathbf{s}}_+^{\mathrm{T}} + r_{pp}\hat{\mathbf{p}}_-\hat{\mathbf{p}}_+^{\mathrm{T}} + r_{ps}\hat{\mathbf{s}}_-\hat{\mathbf{p}}_+^{\mathrm{T}}].$$
(A2)

In Eq. (A1) and Eq. (A2), $k_{0z} = (k_0^2 - k_\parallel^2)^{1/2}$ is the out-of-plane component of the wavevector and $r_{ij}$ is the reflection coefficient of incident $i$-polarization ($s$- or $p$-) to reflected $j$-polarization ($s$- or $p$-). A 4×4 transfer matrix method is used for calculating reflection coefficients from an anisotropic multilayer [46]. $\hat{\mathbf{s}}_\pm$ and $\hat{\mathbf{p}}_\pm$ are the unit vectors for $s$- and $p$-polarization propagating along $\pm z$ directions:

$$\hat{\mathbf{s}}_\pm = \begin{bmatrix} -\sin\phi \\ \cos\phi \\ 0 \end{bmatrix}; \hat{\mathbf{p}}_\pm = k_0^{-1} \begin{bmatrix} \pm k_{0z}\cos\phi \\ \pm k_{0z}\sin\phi \\ -k_\parallel \end{bmatrix} \quad (A3)$$

Aforementioned, in order to satisfy the general equilibrium, the dyadic Green's function includes both the virtual vacuum emission and the structure's emission. To separate these two and get the emission from the structure only, the vacuum contribution is derived below, following the results from Ref. [23,25,35]. The propagating waves for $s$- and $p$-polarization are defined in the k-space as:

$$\mathbf{E}_{\mathrm{vac}}^s(z, \mathbf{k}_\parallel, \omega) = a_s[e^{ik_{0z}z}\hat{\mathbf{s}}_+ $$
$$+ e^{-ik_{0z}z}(r_{ss}\hat{\mathbf{s}}_- + r_{sp}\hat{\mathbf{p}}_-)],$$
$$\mathbf{E}_{\mathrm{vac}}^p(z, \mathbf{k}_\parallel, \omega) = a_p[e^{ik_{0z}z}\hat{\mathbf{p}}_+ $$
$$+ e^{-ik_{0z}z}(r_{ps}\hat{\mathbf{s}}_- + r_{pp}\hat{\mathbf{p}}_-)],$$
(A4a)

and, similarly, for the corresponding magnetic field, the expressions are given by:

$$\mathbf{H}^s_{\text{vac}}(z, \mathbf{k}_\parallel, \omega) = \frac{a_p}{\eta_0}[e^{ik_{0z}z}\hat{\mathbf{s}}_+ + e^{-ik_{0z}z}(r_{pp}\hat{\mathbf{s}}_- + r_{ps}\hat{\mathbf{p}}_-)],$$

$$\mathbf{H}^p_{\text{vac}}(z, \mathbf{k}_\parallel, \omega) = \frac{a_s}{\eta_0}[-e^{ik_{0z}z}\hat{\mathbf{p}}_+ + e^{-ik_{0z}z}(-r_{ss}\hat{\mathbf{p}}_- + r_{sp}\hat{\mathbf{s}}_-)], \quad (A4b)$$

where $\eta_0$ is the characteristic impedance of vacuum, and $a_{s(p)}$ is the amplitude of an s- or p-polarized wave, whose autocorrelation is defined through the amplitude of the field emitted from a blackbody:

$$\langle a_i a_j^* \rangle = \delta_{ij}\delta(\mathbf{k}_\parallel - \mathbf{k}'_\parallel)4\pi\frac{\mu_0\omega}{k_{0z}}\Theta(\omega, T). \quad (A5)$$

Note that in Eq. (A4), without changing the definition of the direction of polarization, $\mathbf{E}^s_{\text{vac}}$ corresponds to $\mathbf{H}^p_{\text{vac}}$, and $\mathbf{E}^p_{\text{vac}}$ corresponds to $\mathbf{H}^s_{\text{vac}}$, as they are perpendicular to each other. Therefore, the vacuum field correlation can be written as:

$$\langle \mathbf{E}_{\text{vac}}(\mathbf{r}_1, \omega) \otimes \mathbf{H}^*_{\text{vac}}(\mathbf{r}_2, \omega) \rangle = \frac{1}{(2\pi)^2}\iint \langle \mathbf{E}_{\text{vac}}(\mathbf{r}_1, \omega) \otimes \mathbf{H}^*_{\text{vac}}(\mathbf{r}_2, \omega) \rangle e^{i\mathbf{k}_\parallel \cdot (\mathbf{R}_1 - \mathbf{R}_2)} d^2\mathbf{k}_\parallel$$

$$= \frac{1}{(2\pi)^2}\iint \left(\langle \mathbf{E}^s_{\text{vac}}(\mathbf{r}_1, \omega)\mathbf{H}^{p\dagger}_{\text{vac}}(\mathbf{r}_2, \omega)\rangle + \langle \mathbf{E}^p_{\text{vac}}(\mathbf{r}_1, \omega)\mathbf{H}^{s\dagger}_{\text{vac}}(\mathbf{r}_2, \omega)\rangle\right) e^{i\mathbf{k}_\parallel \cdot (\mathbf{R}_1 - \mathbf{R}_2)} d^2\mathbf{k}_\parallel, \quad (A6)$$

and using Eq. (A4), the vacuum contribution can be obtained.

### Appendix B: Full expression of the coherence matrix

In Section 2, Eq. (11) contains a curl operator, which only acts on one of the position vectors, $\mathbf{r}_2$. To show how it can be simplified, a further elaboration is shown here based on the approach in Ref. [25].

First, $\bar{\bar{g}}_{12}$ in Eq. (A1) and $\bar{\bar{g}}_{21}$ in Eq. (A2) are the k-space Fourier transform of $\bar{\bar{G}}$ in Eq. (11). Therefore, the curl operator only acts on exponential terms. The matrix form of curl multiplying $\bar{\bar{\alpha}} = \bar{\bar{g}}_{12}e^{i\mathbf{k}_\parallel \cdot (\mathbf{R}_1 - \mathbf{R}_2)}$, for example, is shown below:

$$\nabla_{\mathbf{r}_2} \times \bar{\bar{\alpha}} = \begin{bmatrix} 0 & -\partial z & \partial y \\ \partial z & 0 & -\partial x \\ -\partial y & \partial x & 0 \end{bmatrix}\bar{\bar{\alpha}}$$

$$= i\begin{bmatrix} 0 & k_{0z} & -k_y \\ -k_{0z} & 0 & k_x \\ k_y & -k_x & 0 \end{bmatrix}\bar{\bar{\alpha}} = i\bar{\bar{c}}_1\bar{\bar{\alpha}}, \quad (B1)$$

where $\partial x$ implies the differentiation operator $\partial/\partial x$. In addition, for $\bar{\bar{\beta}} = \bar{\bar{g}}^\dagger_{21}e^{i\mathbf{k}_\parallel \cdot (\mathbf{R}_1 - \mathbf{R}_2)}$, define another matrix for the curl operator as follows:

$$\nabla_{\mathbf{r}_2} \times \bar{\bar{\beta}} = i\begin{bmatrix} 0 & -k^*_{0z} & -k_y \\ k^*_{0z} & 0 & k_x \\ k_y & -k_x & 0 \end{bmatrix}\bar{\bar{\beta}} = i\bar{\bar{c}}_2\bar{\bar{\beta}} \quad (B2)$$

Note that the complex conjugate is taken for $k_{0z}$ only, since $k_x$ and $k_y$ are real. Therefore, the full expression of the field correlation can be derived as follows:

$$\langle \mathbf{E}_{\text{ge}}(\mathbf{r}_1, \omega) \otimes \mathbf{H}^*_{\text{ge}}(\mathbf{r}_2, \omega) \rangle = \frac{4i}{\pi}\Theta(\omega, T)\frac{\bar{\bar{G}}_{12} - \bar{\bar{G}}^\dagger_{21}}{2i}(\nabla_{\mathbf{r}_2} \times)^T$$

$$= \frac{4i\Theta}{\pi}\iint \frac{k_\parallel dk_\parallel dk_\phi}{2i(2\pi)^2}\left[e^{i\mathbf{k}_\parallel \cdot (\mathbf{R}_1 - \mathbf{R}_2)}\bar{\bar{g}}_{12} - \left(\bar{\bar{g}}_{21}e^{i\mathbf{k}_\parallel \cdot (\mathbf{R}_2 - \mathbf{R}_1)}\right)^\dagger\right](\nabla_{\mathbf{r}_2} \times)^T$$

$$= \frac{4i\Theta}{\pi}\iint \frac{k_\parallel dk_\parallel dk_\phi}{2i(2\pi)^2}(\nabla_{\mathbf{r}_2} \times)^T\left[e^{i\mathbf{k}_\parallel \cdot (\mathbf{R}_1 - \mathbf{R}_2)}(\bar{\bar{g}}_{12} - \bar{\bar{g}}^\dagger_{21})\right]. \quad (B3)$$

$k_\phi$ is the azimuth angle of wavevector $\mathbf{k}_\parallel$ in cylindrical coordinates of k-space. For $\bar{\bar{g}}_{12}$ and $\bar{\bar{g}}^\dagger_{21}$, the corresponding curl matrices are $i\bar{\bar{c}}_1$ and $i\bar{\bar{c}}^T_2$. Thus, replacing the curl operators with corresponding matrices and writing Eq. (B3) in cylindrical coordinates leads to:

$$\langle \mathbf{E}_{\text{ge}}(\mathbf{r}_1, \omega) \otimes \mathbf{H}^*_{\text{ge}}(\mathbf{r}_2, \omega) \rangle = \frac{1}{(2\pi)^2}\iint k_\parallel dk_\parallel dk_\phi \frac{2i\Theta}{\pi}(\bar{\bar{g}}_{12}\bar{\bar{c}}^T_1 - \bar{\bar{g}}^\dagger_{21}\bar{\bar{c}}^T_2)e^{ik_\parallel R\cos(k_\phi - \phi)} \quad (B4)$$

The electromagnetic field correlation of the structure can therefore be obtained by subtracting Eq. (A6) from Eq. (B4).